\documentclass[aps,pre,reprint, amsmath, amssymb,superscriptaddress]{revtex4-1}
\usepackage[amssymb]{SIunits}
\usepackage{bm}
\usepackage[retainorgcmds]{IEEEtrantools}
\usepackage{graphicx}
\usepackage{mathrsfs}
\usepackage{amsmath}
\usepackage{amssymb}
\usepackage{color}
\usepackage{amsfonts}
\usepackage{nicefrac}
\usepackage{braket}

\newcommand{\da}{^\dagger}
\newcommand{\ud}{\,\mathrm{d}}
\newcommand{\co}{(Color Online)}

\DeclareMathOperator{\tr}{tr}

\newcommand{\aak}{a_k^\dagger a_k}
\newcommand{\bbk}{b_k^\dagger b_k}
\newcommand{\abk}{a_k^\dagger b_k}
\newcommand{\bak}{b_k^\dagger a_k}

\newcommand{\nsk}{\eta_{k,\sigma} ^\dagger \eta_{k,\sigma}}	
\newcommand{\nak}{\bar{n}_{A,k}}
\newcommand{\nbk}{\bar{n}_{B,k}}
\newcommand{\hak}{\bar{h}_{A,k}}
\newcommand{\hbk}{\bar{h}_{B,k}}

\begin{document}

\title{ Thermalization and entropy production of a 1D bipartite fermionic lattice under influence of a dephasing noise}
\date{\today}
\author{Wellington L. Ribeiro}
\affiliation{Universidade Federal do ABC,  09210-580 Santo Andr\'e, Brazil}
\email{wlribeiro@ufabc.edu.br}

\begin{abstract}
In this paper we study entropy production and transport of heat and matter   in an 1D bipartite fermionic chain. 
Moreover, we input a phenomenological dephasing noise in order to show its effect  on the dynamics of the system. Specially, it is shown the importance of such noise  in the emergence of the entropy production.   For the fluxes, analytical solutions are obtained using both Boltzmann and Fermi-Dirac distribuition.
Finally, we show that a particular fluctuation theorem for energy and matter exchange is obeyed even in the presence of such noise. 

\end{abstract}
\maketitle{}

\section{\label{sec:int}Introduction}

Irreversible thermodynamics is usually referred as an extension of the equilibrium thermodynamics \cite{callen}.  Its bases are the same fundamental postulates from equilibrium thermodynamics, added by the time reversal simmetry of the physical laws. It is the responsible of studying rates and fluxes in a thermodynamical processes, without the need of ``extremely slow" dynamics, as occurs in the equilibrium thermodynamics.
 
For markovian evolutions, the study of fluxes ($J_k$) is based on Onsager Reciprocity Theory (ORT) \cite{onsager1,onsager2}. Its foundations lie at the analysis of generalized forces, which  drives  the processes,  called as \emph{affinities} ($\mathcal{F}_j$), and their responses. For instance, the affinity related to the thermal equilibrium of a bipartite system is the difference of the inverse of the temperatures \citep{callen}. In this sense, the important Fourier's and Fick's laws can be seen as particular cases in which there is no matter and heat flows, respectively.

Notwithstanding, in the standard thermodynamics formulation the quantities are assumed to be deterministic, not prone to fluctuations. That is a consequence of the large number of particles in macroscopic systems, which make the fluctuations negligibly small compared to the average of some thermodynamic quantity \cite{salinas1}.
For a small number of particles these fluctuations become important and quantities such as heat and work must be treated as random variables. One of the most important consequence of the probabilistic approach is the arising of \emph{fluctuation theorems}. These equalities relate forward and time-reversed probabilities in nonequilibrium phenomena.   Quantum and classical fluctuation theorems have been studied  \cite{crooks, jarzynski, landi2016, saito, zon,ciliberto, ciliberto2} and verified \cite{collin, serra}  for different systems in the last two decades.  A very important verification of these probabilistic features  was shown in Ref.\cite{collin}, where the authors studied the work necessary to fold and unfold a RNA molecule. As a result, they find that due to thermal fluctuations, the work required may changes each time the experiment is repeated.

Besides, when the systems  are sufficiently  small or with few particles, classical mechanics is not sufficient anymore. In this sense, it is important and  useful to formulate a thermodynamics with foundations based in quantum mechanics, which is known as \emph{quantum thermodynamics}(see e.g. \cite{horo,kosloff,vinjanampathy}). 
Many efforts have been done in quantum thermodynamics. Among  the many studies there are applications  electronic circuits \cite{pekola2015}, refrigeration of a gas using quantum dots \cite{prance}, heat flow across a single electronic channel \cite{jezouin} and  formulations involving spins and harmonic oscillators  \cite{ohdissi,ohlutz, anichain}.

In addition to the previous examples, transport properties have been of interest in the last years due to the rich range of many-body phenomena that can be studied, as fluxes of energy and particles, nonequilibrium steady states and thermalization,  for instance \cite{vznidarivc,prosen2010,kaufman2016,medina}.  

In the study of fluxes, one-dimensional systems have  already been  studied classically \cite{rieder,stefano,narayan} and quantically \cite{landi2016}. Still in classical mechanics, conservative noises proved to be useful in the process of finding Fourier's law for one-dimensional chains \cite{landi2013}. It motivates us to study the possible effects of energy-conserving noises in quantum systems,  in the context of one-dimensional systems. 

In this paper we investigate in detail  the role of an energy-conserving noise in the evolution of an 1D bipartite fermionic chain, whose halves are prepared in equilibrium with different thermal and matter reservoirs. The heat and matter exchanges are studied through the Onsager's coefficients. The entropy evolution and the dependence of the openness of the system and the entropy production is derived. Finally, we show that a fluctuation theorem for matter and energy exchange is also obeyed. 

The paper is organized as follows: in section \ref{model} the model and basic assumptions  are presented. In  \ref{heatsec}, the results of fluxes of heat and particles are shown. A brief comparison between closed and open evolutions are also made. In section \ref{entrosec}, the evolution of the entropy and mutual information are shown.  In section \ref{flucsec} we obtain the fluctuation theorem  for heat and matter considering the open system evolution. In  \ref{conclusec} the conclusions of the behaviour of the system is made. In the appendix it is shown how to obtain analytical solutions when we use both Fermi-Dirac and Boltzmann's statistics. In particular, in the case of Boltzmann's statistics, a function $\omega_\nu (x,y)$ is derived in order to simplify  obtaining the thermodynamic quantities.   Although it was only used in calculation of fluxes, it is general and can be used to compute entropy and mutual information.      

\section{The model} \label{model}

\subsection{Closed properties of the system}
\begin{figure}[!h]
\centering
\includegraphics[scale=0.5]{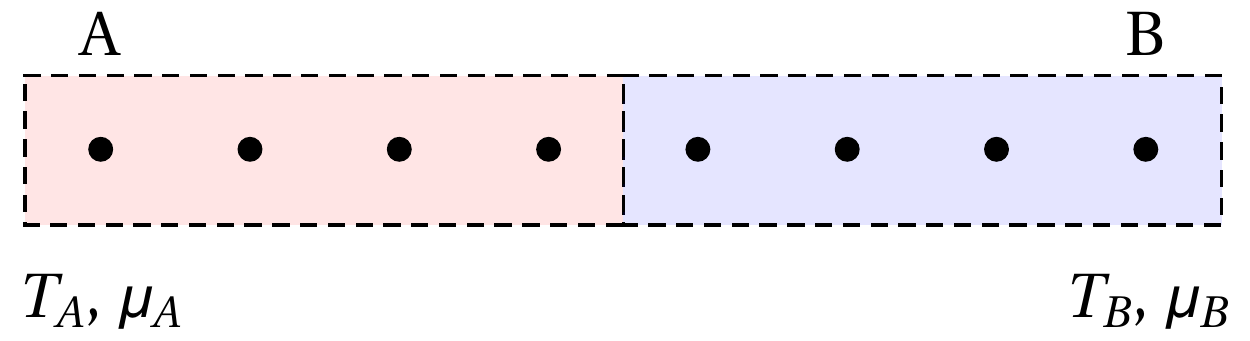}
\caption{(Color Online) Representation of the bipartite lattice, in which each half is prepared in  equilibrium with a different reservoir of energy and matter.}
\label{chain}
\end{figure}

As a start point, let us  consider a fermionic 1D bipartite lattice, composed by $2N$ sites. The system is divided into halves that will be called  as $A$ and $B$. Each subsystem is prepared in thermal equilibrium with a heat and matter reservoir, defined by temperatures $T_A$  and $T_B$ and chemical potentials $\mu_A$  and $\mu_B$ respectively, as represented in Fig.\ref{chain}.

 The system is modeled in the second quantization language, in the  position representation, through  creation (annihilation) operators $c_i \da$ ($c_i$), in which each site $i$ is connected to the nearest-neighbors according to the  tight-binding Hamiltonian
\begin{equation}
H = - \sum_{i=1}^{2N -1} \alpha_i (c_{i+ 1}\da c_i + c_i\da c_{i+1}),
\end{equation} 
in which $\alpha_i$ are the coupling constants between sites and $N$ is the number of sites in each half. As we are dealing with a bipartite system, it is convenient to define a new set of operators for each subsystem. We define   $a_i = c_i$ for $A$  and   $b_i = c_{i+N}$ for $B$, taking $i = 1,2,3,\cdots, N$. Moreover, we choose all coupling constants $\alpha _i = 1$ (which sets the energy scale) except for $i=N$, in which we choose $\alpha_N = g_0$. In order to use the weak-coupling approximation, we will take $g_0 \ll 1$. 

Given the present set of coupling constants,  we divide the Hamiltonian as	 
\begin{equation}
H = H_A + H_B + V,
\end{equation}
where $H_X$, with $X =A,B$, are the Hamiltonians of each subsystem and $V$ describes the coupling between the subsystems, such that
\begin{eqnarray}
H_A &=& - \sum_{i=1}^{N-1} (a_{i+1}\da a_i + a_{i}\da a_{i+1}),\\
H_B &=& - \sum_{i=1}^{N-1} (b_{i+1}\da b_i + b_{i}\da b_{i+1}),\\
V &=& -g_0(a_N \da b_1 + b_1\da a_N).
\end{eqnarray} 
Due to the open boundary conditions and the homogeneity of each half, we diagonalize the Hamiltonians $H_X$ computing the momenta representation, via Fourier's sine transformation, given by
\begin{equation}
a_i = \sum_k S_{i,k} a_k, \qquad  b_i = \sum_k S_{i,k} b_k, 
\end{equation}
in which $S_{i,k} = \sqrt{\frac{2}{N+1}}\sin(ki)$. Besides,  $k \in [0, \pi]$ such that $k = \frac{\pi }{N+1}, \frac{2\pi}{N+1},\cdots, \frac{N \pi}{N+1}$, always with steps of $\frac{\pi}{N+1}$. With this transformation, we  straightforwardly obtain
\begin{equation}
H_A = \sum_k \epsilon_k \aak \quad \text{and} \quad H_B = \sum_k \epsilon_k \bbk, 
\end{equation} 
with $\epsilon_k = - 2 \cos(k)$.

The interaction between $A$ and $B$, in the momenta representation, is given by
\begin{equation}
V =-\frac{2g_0}{N+1} \sum_{k,q} \sin(Nk)\sin(q)(a_k\da b_q + b_q\da a_k),
\end{equation}
which means that all modes are connected. Even so, if we deal with the problem in the interaction picture, we see that each term $a_k\da b_q$ are followed by an exponential $e^{i(\epsilon_k - \epsilon_q)t}$. For the limits of small $g_0$, we can use the rotating wave approximation (RWA), since terms with $k \neq q$ will oscillates rapidly and will have small contributions to the dynamics compared to the other terms ($k = q$). Thus, using this approaximation, $V$ is given by 
\begin{equation}
V \approx \sum_k g_k (a_k\da b_k + b_k\da a_k),
\end{equation} 
in which we defined $ g_k = -\frac{2g_0}{N+1}\sin(Nk)\sin(k)\equiv g\sin(Nk)\sin(k) $.

Besides, the Hamiltonian for the full system can now be written as $ \displaystyle H = \sum_k H_k$, where
\begin{equation} \label{hamilk}
H_k = \epsilon_k (\aak + \bbk) + g_k (a_k\da b_k + b_k\da a_k).
\end{equation} 

\begin{figure}[!h]
\centering
\includegraphics[scale=0.5]{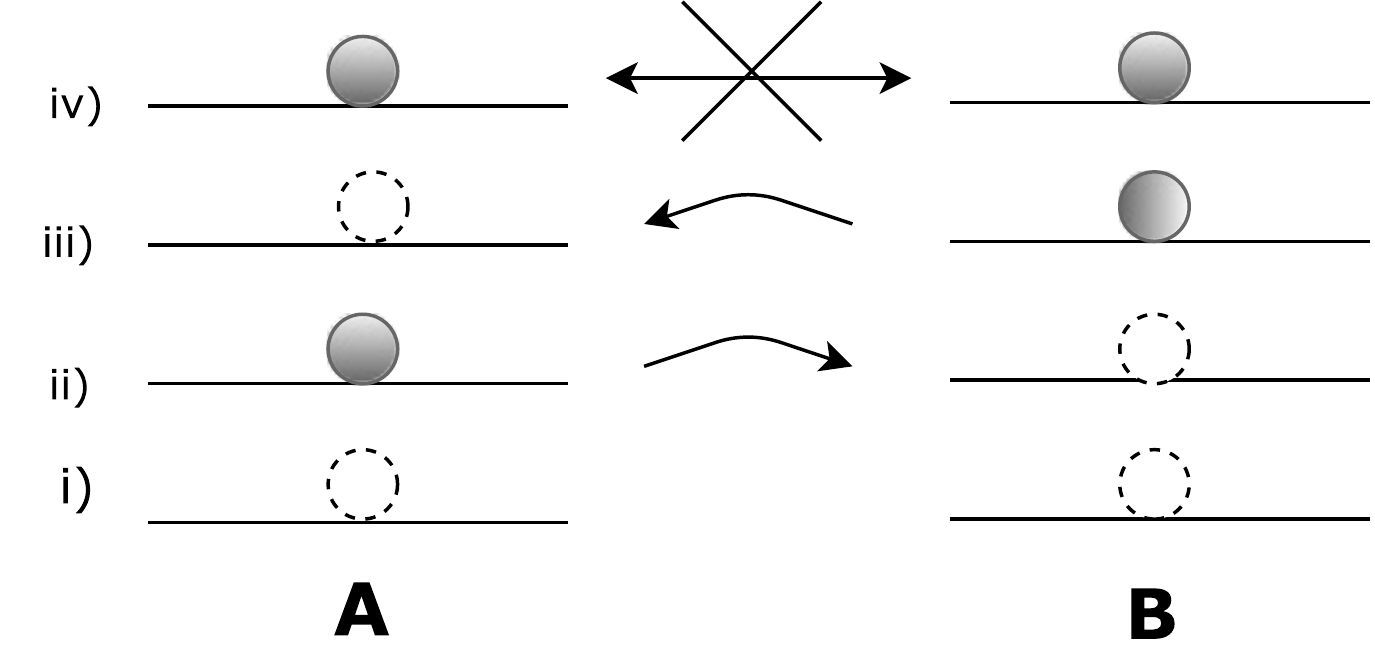}
\caption{Representation of the possible dynamics for a given mode. 
i) There are no particles in the chain A, as well as in chain B,
ii) Chain A with one particle and B with no particles. Exchange is allowed,
iii) Chain B with one particle and A with no particles. Exchange is allowed,
iv) There is one particle in each chain. There are no exchanges.}
\label{Xchanges}
\end{figure}

 Therefore, the RWA allows us to uncouple the modes  and, as we will see later, the dynamics of the system. The Fig.\ref{Xchanges} represents the possible dynamics for each (uncoupled) mode, already considering Pauli exclusion principle.
Lastly, in order to finish the diagonalization of the Hamiltonian, we define a new set of operators
\begin{equation}
\eta_{k,\sigma}= \frac{a_k + \sigma b_k}{\sqrt{2}},
\end{equation}
where $\sigma = \pm 1$ and find
\begin{equation} \label{hamileta}
H = \sum_{k,\sigma} (\epsilon_k + \sigma g_k) \nsk.
\end{equation}

It is worth to comment that both shapes of the Hamiltonian, given by eq.(\ref{hamilk}) and eq.(\ref{hamileta})  have their advantages. The dynamics of the system is clearer understood by working with the operators $a_k$ and $b_k$. On the other hand, the construction of the dephasing noise, as well as computing the dynamics, are easier in the basis of $\eta_{k,\sigma}$.

\subsection{Dephasing noise and density matrix}

Now we turn our attention to the  bases of the evolution of the system. First, we consider that the system evolves under the influence of an environment. In order to study the role of decoherence over the system, we will not consider dissipation effects.  To achieve this goal, only a phenomenological dephasing noise will be used.

Once the Hamiltonian is factorized into modes, it is sufficient to compute the evolution of the reduced density matrix for each one. Moreover, we consider a Markovian evolution, and then the dynamics of the system will be computed by using Lindblad master equation 
\begin{equation} \label{lindblad}
\dot{\rho}_k(t)= -i[H_k,\rho_k(t)] + \lambda D(\rho_k(t)),
\end{equation}
in which
\begin{equation}
D(\rho_k(t)) = \sum_\sigma  L_{\sigma,k} \tilde{\rho}_{k} L_{\sigma,k}\da - \tfrac{1}{2}\{L_{\sigma,k}\da L_{\sigma,k}, \rho_k \},
\end{equation}
where $L_{k,\sigma}$ are the Lindblad generators.

 As soon as  the dephasing noise conserves energy, we can build it using the condition $[H,L_{k,\sigma}]=0$. For each subspace defined by $k,\sigma$ we choose $L_{k,\sigma} = \nsk $, such that the non-unitary part of the dynamics $\tilde{D}(\tilde{\rho}_{k,\sigma})$ is given by
\begin{equation}
\tilde{D}(\tilde{\rho}_{k}) = \sum_{\sigma} \nsk \tilde{\rho}_{k} \nsk - \frac{1}{2}\{(\nsk)^2, \tilde{\rho}_k \},
\end{equation}
in which we have adopted the notation $\tilde{O}$ to make explicit the cases in which an  operator $O$ is in the basis of $\eta_{k,\sigma}$.
 
As mentioned in \ref{model}, we prepare each half in a different thermal state. In the time $t=0$ we connect them and let the total system $A+B$ evolve. The dynamic of the system is obtained by solving  Lindblad master equation in the basis $\{ \ket{0}, a_k\da \ket{0}, b_k\da \ket{0}, a_k\da b_k\da \ket{0} \}$, we find 
\begin{widetext}
\begin{equation} \label{densitymatrix}
\rho_k (t) = 
\begin{pmatrix}
\hak \hbk &0&0&0 \\
0 &\braket{\aak}_t - \nak \nbk& \braket{\bak}_t& 0 \\
0 & \braket{\abk}_t & \braket{\bbk}_t - \nak \nbk  &0 \\
0&0&0& \nak \nbk
\end{pmatrix}
\end{equation}
where $\bar{n}_{X,k} = (e^{\beta_X(\epsilon_k - \mu _X)}+1)^{-1}$ is the Fermi-Dirac distribution, $\bar{h}_{X,k}:= 1 - \bar{n}_{X,k}$. We already introduced the  expectation values
\begin{eqnarray}
\braket{\aak}_t &=& \frac{(\nak + \nbk)}{2} + \frac{(\nak - \nbk)}{2} e^{- \lambda t} \cos (2 g_k t) \label{Amean} \\
\braket{\bbk}_t &=& \frac{(\nak + \nbk)}{2} - \frac{(\nak - \nbk)}{2} e^{- \lambda t} \cos (2 g_k t) \label{Bmean} \\
\braket{\abk}_t &=& \frac{i}{2}(\nak - \nbk) e^{- \lambda t} \sin (2 g_k t). \label{ABmean}
\end{eqnarray} 
\end{widetext}

\section{Heat and particle fluxes} \label{heatsec}

In this section we study heat and paticle fluxes through the new mathematical quantities $\bar{N}$ and $\bar{E}$, defined, for instance, by using the averages of occupations. 

In this work, in order to avoid some finite features, we will consider $N \to \infty$. Moreover, as soon as the physical quantities would diverges for this limit, they are  defined and computed \emph{per site}. As a particular interest of this work, the subsystems are prepared slightly out-of-equilibrium to each other, so we can compute the evolution only using linear terms of the affinity. In other words, we will use $T_A = T + \frac{\delta T}{2}$ and $\mu_A = \mu + \frac{\delta \mu}{2}$ for the half $A$ and  $T_B = T - \frac{\delta T}{2}$ and $\mu_B = \mu - \frac{\delta \mu}{2}$ for the half $B$,  in such way that $\frac{\delta T}{T} \ll 1$ and $\frac{\delta \mu}{|\mu|} \ll 1$. These assumptions are justified by remembering that they may lead  the well-known Fourier's law (for heat conduction), Fick's law (for matter conduction) and mainly thermoelectrical effects.

Through the expactation values given in  eq.(\ref{Amean}-\ref{Bmean}) we obtain both the average number of particles and energy. As the particles are independent, such quantities for $A$, for example,  are given by
\begin{equation}
\braket{N_A} = \sum_k \braket{\aak} \quad \text{and} \quad \braket{H_A} = \sum_k \epsilon_k \braket{\aak}.
\end{equation}  	
For $B$, this idea is completely analogue. The change of the number of particles of one subsystem may be studied through the quantity
\begin{equation}
\mathcal{J_N}(t) = \frac{\braket{N_A(t)}-\braket{N_A(0)}}{N} 
\end{equation}  
that is how many  particles (in average), per site, flowed to  $A$ until the time $t$. For the sake of simplicity,  such quantity will be referred by \emph{particle flux}, albeit this is not the time derivative  of $\braket{N_A(t)}$. An analogue case can be made for the heat. Through the first law of thermodynamics and the averages of energy and particles, we define the total  heat flowed to $A$, per site, until the time $t$ by
\begin{equation}
\mathcal{J_Q}(t)= \frac{(\braket{H_A(t)}-\braket{H_A(0)})- \mu (\braket{N_A(t)})- \braket{N_A(0)})}{N}, 
\end{equation}
in which it will be called by \emph{heat flux}. 

In order to compute these quantities, we use two main informations: i) as we are taking $N\to\infty$, which allows us the replacement  $\sum_k\to\tfrac{N}{\pi} \int \ud k$, ii) due to the systems are prepared slightly out of equilibrium, the fluxes can be described in terms of Onsager (linear) coefficients as follows:
\begin{equation}
\begin{pmatrix}
\mathcal{J_N}(t) \\
\mathcal{J_Q}(t)
\end{pmatrix} = \begin{pmatrix}
\frac{T}{2} \frac{\partial \bar{N}}{\partial \mu} & \frac{T^2}{2}\frac{\partial \bar{N}}{\partial T} \\
\frac{T}{2} \frac{\partial \bar{Q}}{\partial \mu} & \frac{T^2}{2}\frac{\partial \bar{Q}}{\partial T} 
\end{pmatrix} \begin{pmatrix}
\frac{\delta \mu} {T} \\
 \frac{\delta T}{T^2}
\end{pmatrix}.
\end{equation}

The quantity $\bar{Q} := \bar{E}- \mu \bar{N}$ in which
\begin{eqnarray}
\bar{N} &=& \frac{1}{\pi} \int_0^\pi \bar{n}(\mu,T,k) (e^{-\lambda t}\cos(2g_kt)-1) \ud k \\
\bar{E} &=& \frac{1}{\pi} \int_0^\pi \bar{n}(\mu,T,k) \epsilon_k (e^{-\lambda t}\cos(2g_kt)-1) \ud k. 
\end{eqnarray} 
Despite of $\bar{N}$ and $\bar{E}$ were only mathematically defined, they contain all needed information about the evolution. Such functions will be solved using Fermi-Dirac and Boltzmann statistics in  Appendix \ref{analytical}.  

Fermi-Dirac distribution is more general (or fundamental) than Boltzmann distribution. But in insulators and semiconductors, the last level occupied (at $T= 0K$) belongs to the valence band. For a finite temperature, some electrons of the material may be thermally excited. In such cases, the Fermi-Dirac distribution is well approximated by Boltzmann distribuition.

The dynamics depends directly on the occupation of each mode. The replacement $\bar{n}_{FD} = [e^{\beta(\epsilon - \mu)}+1]^{-1} \to \bar{n}_B = e^{-\beta(\epsilon - \mu)}$ naturally provides a theoritical error, which depends on temperature, chemical potential and the mode that we are analyzing. As soon as we are going to higher energy modes, the Fermi-Dirac gets closer to Boltzmann distribution for a given temperature and chemical potential. Thus, the major error will be related to the mode with the lower energy ($k = 0$). Thus, recovering for this analysis  we are able to estimate how good is our approximation through
\begin{equation} \label{erroroccupation}
|\bar{n}_{FD} - \bar{n}_{B}|  < 10^ {-m},
\end{equation}  
in which $m$ defines how small is the difference between the both distributions. Moreover, the eq.(\ref{erroroccupation}) gives us  a relation between the temperature and chemical potential for this approximation. Thus, recovering $k_B$ and $\alpha$ for this analysis, we obtain
\begin{equation}\label{muboltz}
\mu < -\frac{m k_B T \ln 10}{2} -2 \alpha. 
\end{equation}

The quantity $E_{gap} \equiv \frac{m k_B T \ln 10}{2} < -2 \alpha - \mu$ may be interpreted as the band gap for a material in which the theoretical error of the approximation is of order $10^{-m}$. For the room temperature ($300$K) and  $m = 10$ we obtain $E_{gap} \approx 0.29\text{eV}$, which is approximated value for the band gap for semicondutors like PbTe and PbSe. If we take $m = 12$, we obtain the energy scale of the band gap of the InAs and PbS. Finally, it is worth to comment that for an error of $10^{-m}$, the occupation $\bar{n}_{B} \approx 10^{-\frac{m}{2}}$, which ensures the validity of the approaximation.

We plot both the equilibrium  and the evolution  for the quantities $\mathcal{J}_{N,\mu} \equiv \frac{T}{2} \frac{\partial \bar{N}}{\partial \mu}$ (adimensional) , $\mathcal{J}_{N,T} \equiv \frac{T^2}{2} \frac{\partial \bar{N}}{\partial T}$ (in units of $\alpha$), $\mathcal{J}_{Q,\mu} \equiv \frac{T}{2} \frac{\partial \bar{Q}}{\partial \mu}$(in units of $\alpha$) and $\mathcal{J}_{Q,T} \equiv \frac{T^2}{2} \frac{\partial \bar{Q}}{\partial T}$(in units of $\alpha^2$). For simplicity, we will name them as fluxes of particles(heat) due to the difference of chemical potential (temperature). 

The following presented results were obtained numerically. The analytical solutions and their validity are shown in appendix \ref{analytical}.
In Fig.\ref{ons1} we have the final fluxes. We observe that such fluxes are simmetric in relation to $\mu =0$. This feature appears due to simmetry of our band of energy  in relation to $\epsilon_k =0$. Moreover, the fluxes are strongly surpressed as soon as $|\mu|$ grows. For the case of $\mu < -2$, it occurs because the average occupation is very low, for both halves. On the other hand, for  $\mu > 2$, all states are much likely  to be occupied and Pauli  exclusion principle prohibits us of having fluxes.

The flux of particles due to the  difference of chemical potential gives us a better understanding of the relation between the chemical potential and the temperature over the fluxes. Two peaks appear on such fluxes and they are near to $\mu = \pm 2$. These values are those in which or one chain begun to be occupied ($\mu = -2$) and the other not or one chain to be fully occupied and the other not ($\mu = 2$). This effect is then smoothed by the temperature, in which for $\mu < -2$ it enhances the probability of a mode to be occupied, while when $\mu > 2$ the temperature decrease the probability of the states to be occupied.  
\begin{figure}[!h]
\includegraphics[width = 0.485 \textwidth ]{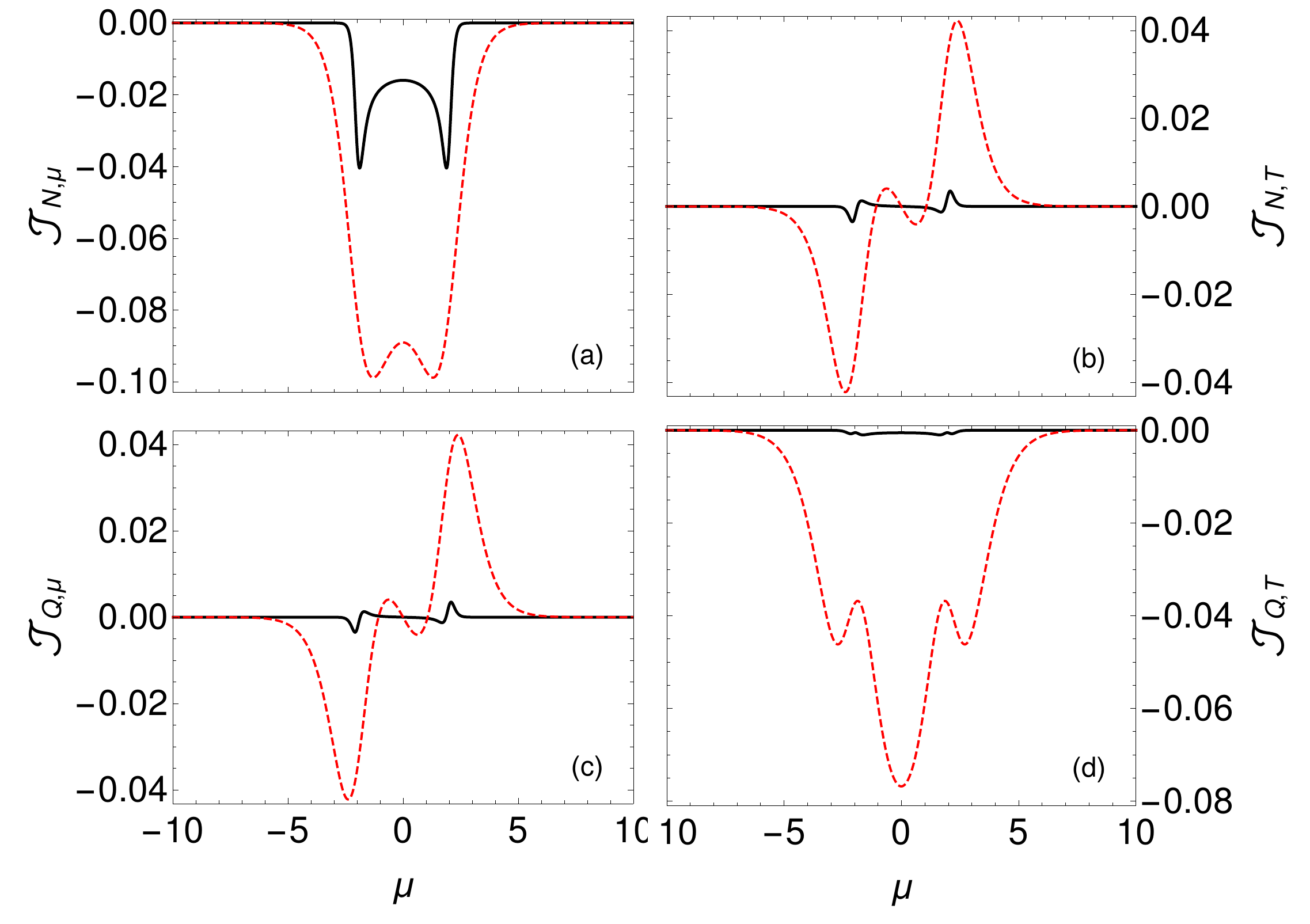}
\caption{\co : Onsager coefficients for $t \to \infty$ (equilibrium) as function of chemical potential in which  $\mu$ and $T$ are in units of $\alpha$. The solid black lines represent the fluxes for $T= 0.1$ and the dashed red lines represent the fluxes for $T = 0.5$. }
\label{ons1}
\end{figure}

\begin{figure}[!h]
\flushright
\includegraphics[width = 0.485 \textwidth ]{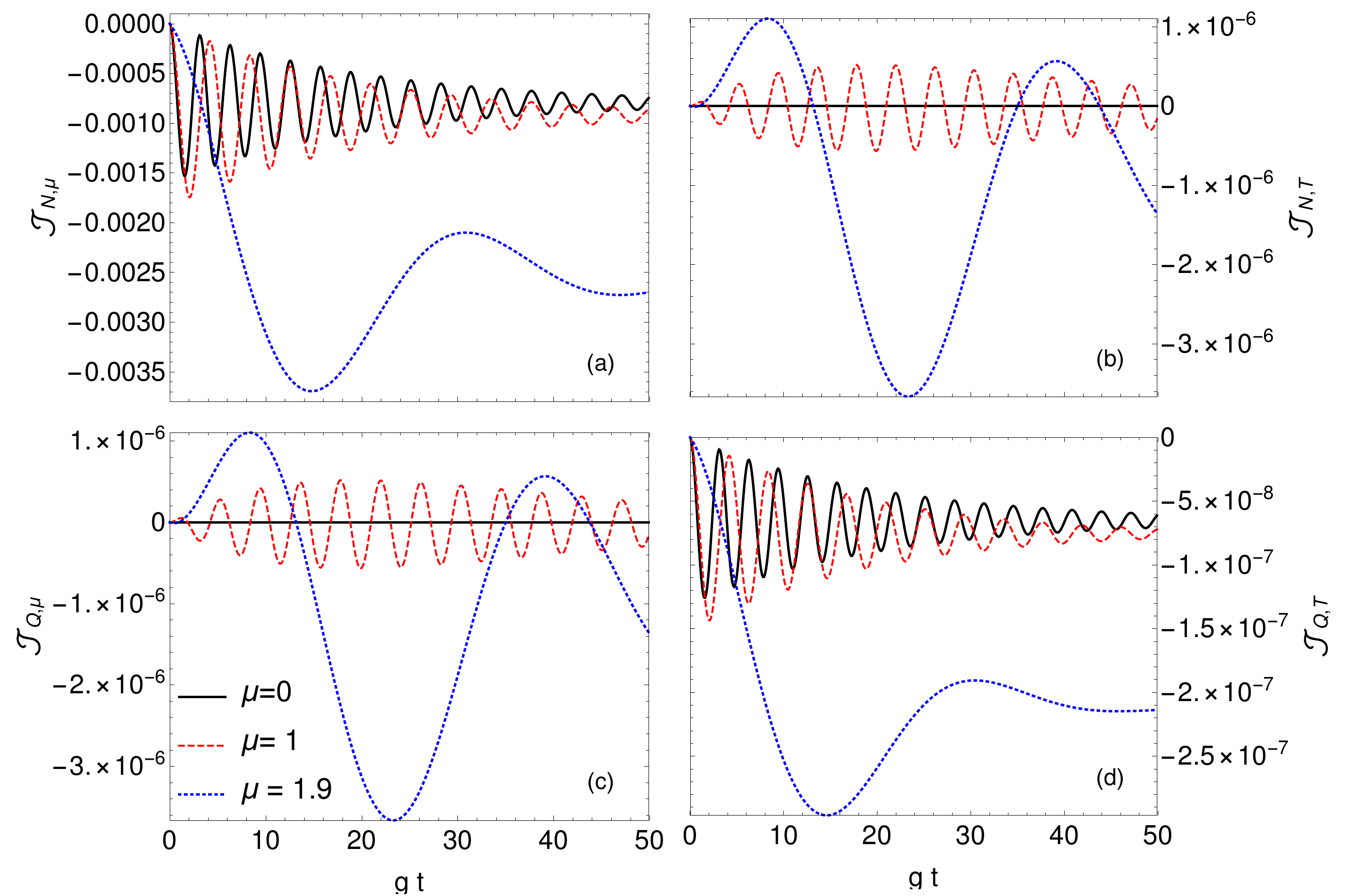}
\caption{\co : The evolution of Onsager coefficients for $T= 0.005$ and $\lambda = 0.05$. In the solid black line we have $\mu =0$, in the red dashed line $\mu =1$ and blue dotted line  $\mu =1.9$.}
\label{onsevo1}
\end{figure}

\begin{figure}[!h]
\flushright
\includegraphics[width = 0.51 \textwidth ]{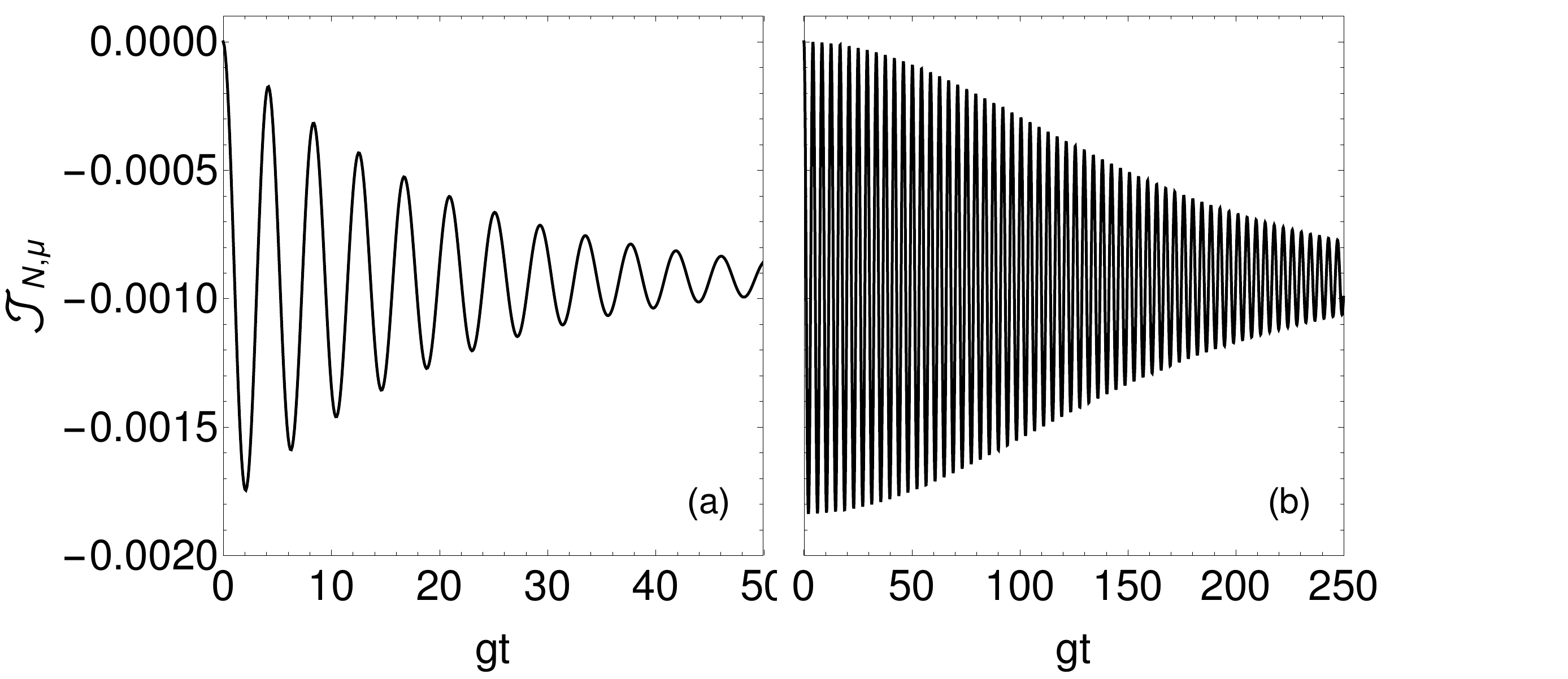}
\caption{\co : Comparative of the evolution of $\mathcal{J}_{N,\mu}$ with (left) and without (right) a dephasing in $T= 0.005$. In (a) we used $\lambda = 0.05$ and in (b) we used $\lambda = 0$.}
\label{onsevo2}
\end{figure}

The evolution of the fluxes are represented in Fig.\ref{onsevo1} and Fig.\ref{onsevo2}. An oscillatory behavior of the fluxes can be  observed. This feature is the result of the multiple exchanges of particles occuring between both halves and repeated  exchanges envolving the same modes. Moreover,  the oscillatory behavior is dependent of the chemical potential. As soon as the chemical potential grows, 	this feature becomes smoother, due to the reduced number of possible exhanges. Finally, as we can see in Fig.\ref{onsevo2}, the dephasing noise does not interfer, as expected, in the intensities of the fluxes or final results, but helps the system to thermalize.

\section{Entropies and mutual Information} \label{entrosec}

In this section we will discuss some bases from entropy theory and then apply to the system of interest.

The concept of entropy plays a central role in statistical mechanics due to its connection with thermodynamics and information theory.
The von Neumann entropy provides an approach to quantum statistical mechanics and thermodynamics and it is defined as
\begin{equation} \label{entropydef}
S(\rho) \equiv - \tr\{\rho  \ln(\rho) \}, \quad k_B = 1 .
\end{equation}

For  a bipartite system, the total (joint) entropy is related to the individual entropies through 
\begin{equation} \label{sinequal}
S(\rho_{AB}) \leq S(\rho_A) + S(\rho_B),
\end{equation}
where the equality holds for $\rho_{AB} =  \rho _A \otimes \rho _B$. The joint entropy measures our total lack of information about a composite system \cite{chuang}.

Equation (\ref{sinequal}) is the starting point for quantifying the degree of correlation between two systems by means of the \emph{mutual information}. In other words, how much correlated they are in a sense that how much knowing about one of these systems reduces the uncertainty about the other system.
For instance, if two systems $A$ and $B$ are independent, then knowing $A$ does not give us any information about $B$ and vice versa, so their mutual information is null. The mutual information as a function of the individual entropies and the total entropy is defined by \cite{zeng}
\begin{equation} \label{mutual}
\mathcal{I} \equiv S(\rho_A) + S(\rho_B) - S(\rho_{AB}).
\end{equation}

For the total entropy, we use Eq.(\ref{densitymatrix}). For each subsystem, we use a reduced density matrix. For instance, for a mode of  $A$ we have
\begin{equation}
\rho_{A_k}(t) = \begin{pmatrix}
1- \braket{\aak}_t & 0 \\
0 & \braket{\aak}_t
\end{pmatrix}.
\end{equation}
Moreover, it is worth to comment that we obtain $\rho_{B_k}(t)$ just by doing $\braket{\aak}_t \to \braket{\bbk}_t$.

As used in the analysis of the fluxes, the subsystems are prepared slightly out-of-equilibrium, such that   $n_{A,k} = n_k + \frac{\delta n_k}{2}$ and $n_{B,k} = n_k - \frac{\delta n_k}{2}$, in which $\delta n_k = \delta T \frac{\partial n_k}{\partial T} + \delta \mu \frac{\partial n_k}{\partial \mu}$. Using Eq.(\ref{entropydef}), expanded  in Taylor series until the second order, we obtain 
\begin{equation}\label{entroa}
S_{A,k}(t) = \sum_{i=0,1,2} S_{k}^{(i)}(\delta n_k)^i
\end{equation}
\begin{equation}\label{entroa}
S_{B,k}(t) = \sum_{i=0,1,2}(-1)^i S_{k}^{(i)}(\delta n_k)^i
\end{equation}
in which
\begin{equation}
S_{k}^{(0)}(t) = -(1-n_k)\ln(1-n_k) -n_k\ln(n_k),
\end{equation}
\begin{equation}
S_{k}^{(1)}(t)=   \frac{1}{2}e^{-\lambda t}\cos(2g_kt) \bigg(\ln(1-n_k)-\ln(n_k)\bigg),
\end{equation}
and
\begin{equation}
S_{k}^{(2)}(t)=\frac{e^{-2 \lambda t}\cos^2(2g_kt)}{8(n_k-1)n_k}.
\end{equation}

As it is possible to note, the entropy for  $B$ (and momentum $k$) is the same of $A$ up to the sign in the first order dependent term (or for any odd power of $\delta n_k$). Thus, once we compute $S_{A,k}(t)$, $S_{B,k}(t)$ is obtained straightfowardly.   

Through the definition, once we expand the entropies, the mutual information for a mode $k$ is given by
\begin{equation} \label{mutualinf}
\mathcal{I}_k(t) = \frac{e^{-2\lambda t}\sin^2(2g_k t)}{4n_k(1-n_k)}\delta n ^2_k .  
\end{equation}
So, having on hands Eq.(\ref{entroa}) and Eq.(\ref{mutualinf}) we are able to obtain the full entropies. Entropy (and as consequence, entropy production) is an additive quantity. Thus, analyze one-mode evolution will gives us the information about the dynamics of the system, without loss of generalization.  
\begin{figure}[!h]
\centering
\includegraphics[scale = 0.57 ]{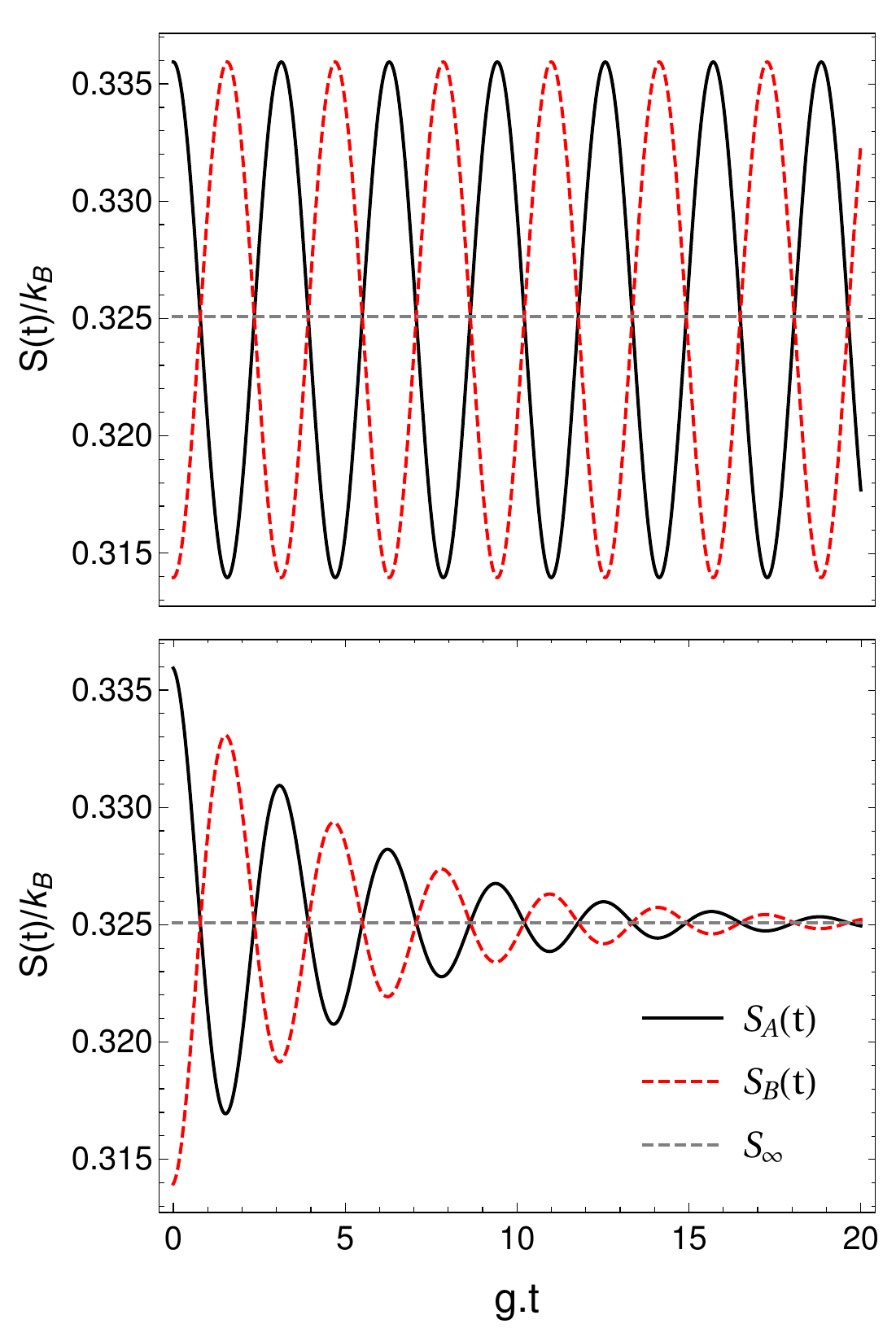}
\caption{\co : Comparative of the evolution of one-mode entropies with (top) and without (bottom) a dephasing. The systems are prepared at $\bar{n}_{eq}= 0.1$, $\delta n =0.01$, $\lambda= 0.2$ (for the bottom panel) and $g=1$. The solid black line represents $S_A(t)$ and the dashed red line represents $S_B(t)$.}
\label{entroevo}
\end{figure}

In Fig.\ref{entroevo} we have a comparative of the entropy evolution for  $A$ and $B$. In this picture we just prepared the modes slightly out of equilibrium by defining the equilibrium  and  initial difference occupations. For one mode analysis, it is not a problem, because temperature and chemical potential (and their differences) only define the initial occupations (and their differences, as well) and not their evolution.    

If we just look the evolution of the system through the Fig.\ref{onsevo2}, we may imagine that the dephasing   is only a influence in the time scale of the thermalization. It is partially true, but looking  the one mode entropies we can see that without dephasing, exchanges of particles will happen in every instant of time, which does not happen in the case with dephasing noise.

Considering the full system, after a sufficient long time, we obtain through the fluxes analysis the same value of the expected from the  standart formulation of thermodynamics. Then, in a first analysis, it looks like that the dephasing is no needed to the thermalization of the system.  Even so, Fig.\ref{entroevo} shows us that the total entropy is constant (without dephasing), going on the other hand of the expected. Thus, the dephasing is a main ingredient in the validity of  the second law of the thermodynamics more than in the thermalization itself.

Through the definition $S_{AB,k}=S_{A,k}(t)+ S_{B,k}(t) - \mathcal{I}_k(t)$,  we can compute the entropy production $\Pi_k(t) \equiv \frac{dS_{AB,k}}{dt} $. Straightfowardly we obtain
\begin{equation}\label{prod}
\Pi_k (t) = \frac{1}{2} \frac{\lambda e^{-2\lambda t}}{(1-n_k)n_k} \delta n ^2_k.
\end{equation}

\begin{figure}[!h]
\flushright
\includegraphics[width = 0.5 \textwidth ]{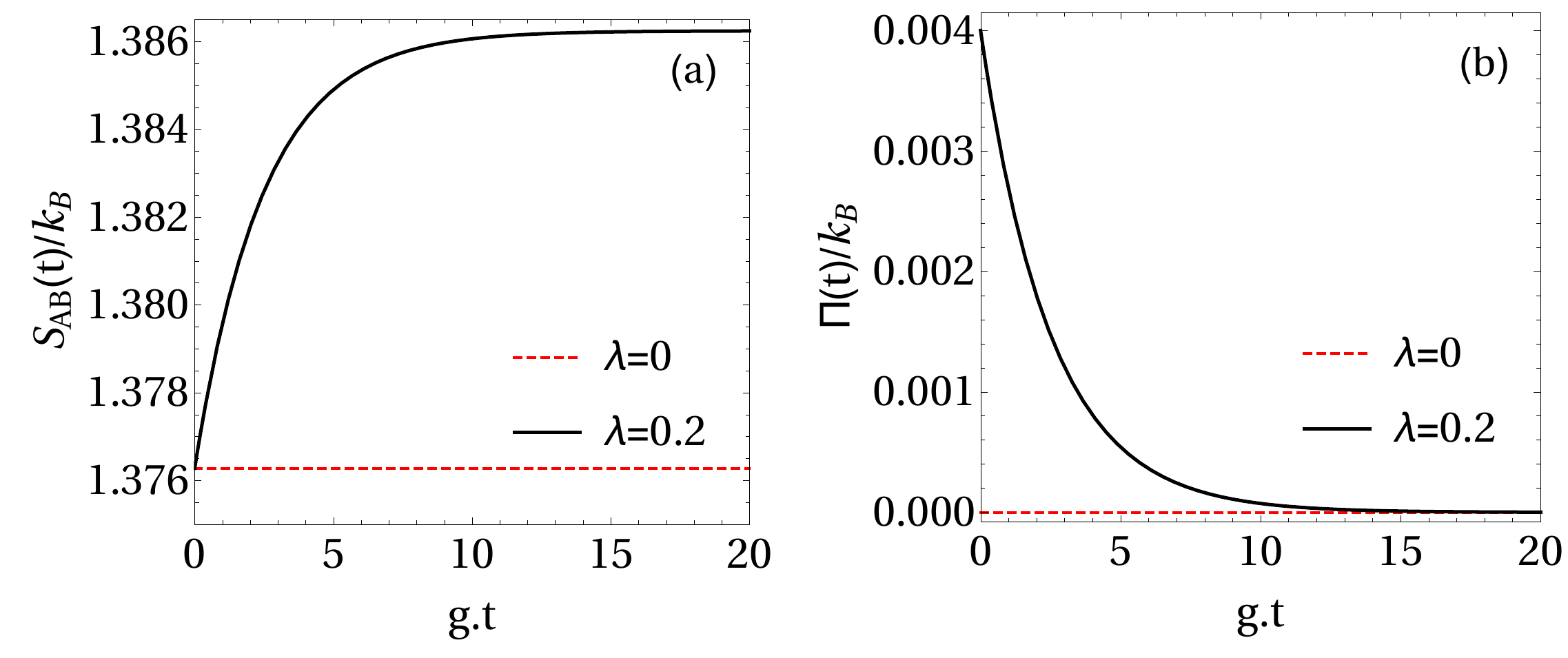}
\caption{\co :(a) One-mode entropy for the composite system with (black line)  and without (red dashed) a dephasing. (b) One-mode entropy production  with (black line)  and without (red dashed) a dephasing. The systems are prepared at $\bar{n}_{eq}= 0.5$, $\delta n =0.1$, $\lambda= 0.2$ (for the case with dephasing) and $g=1$. }
\label{entroprod}
\end{figure}
So, by the Eq.(\ref{prod}) and Fig.\ref{entroprod} we can see that the dephasing noise has in fact a fundamental role in the obtaining of the second law of thermodynamics and even in the connection between the quantum and classical treatments.  
\begin{figure}[!h]
\includegraphics[width = 0.4 \textwidth ]{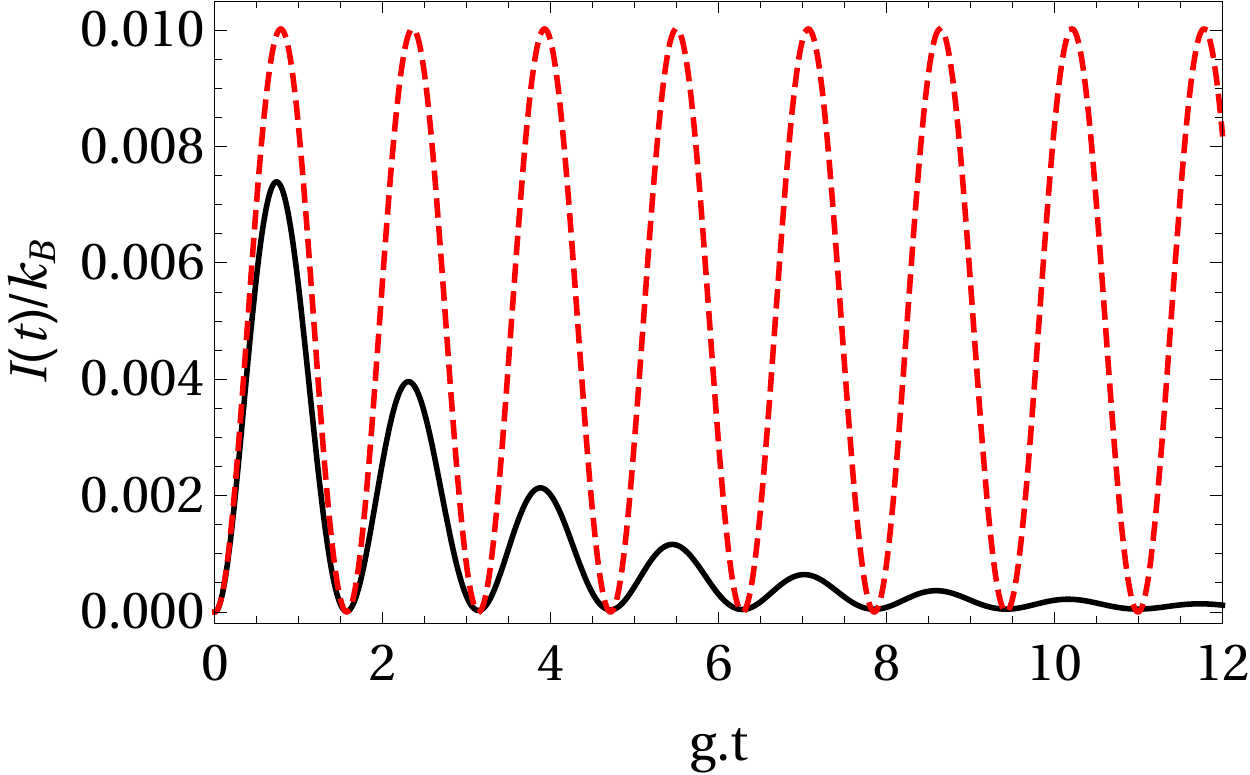}
\caption{\co :(a) One-mode mutual information. The black line represents the evolution with dephasing  and the red dashed one, without  a dephasing.  The systems are prepared at $\bar{n}_{eq}= 0.5$, $\delta n =0.1$, $\lambda= 0.2$ (for the case with dephasing) and $g=1$. }
\label{mutint}
\end{figure}
Finally, in Fig.\ref{mutint} we can see that without the dephasing, the system mantains correlated even for a long time. By a time derivative of Eq.(\ref{mutual}) we can see that the rate of variation of it is directly related to the entropy production. So, as soon as the mutual information fades away, we have an production of entropy.

\section{Fluctuation theorem for energy and matter} \label{flucsec}
In this section we will discuss about the achievement of a fluctuation theorem of matter and energy exchanges. We show that for the open quantum system, the  obtained equallity is the same obtained previously without considering an openness of the system. 

Due to the weak coupling we were able to decouple the different momenta dependence, in other words, the creation of a particle with momentum $k$ in a chain has no influence on the destruction of a particle with momentum $q$ of the another chain and vice versa. So, the processes of exchange of energy are related to the exchange of particles. Thus, if there are no particles in both chains for a certain value of momentum, then we expect that this level will not contribute to the exchange of energy. In the case where a certain momentum $k$ is occupied in both chains, due to the Pauli Exclusion Principle, then we will not have an exchange of energy either. 

For a given momentum, the probability of a system exchange an energy $E_k$ is given by
\begin{equation} \label{probxchange}
\small
P_k(E_k) = \sum_{n,m} \bra{m} \rho_k(t) \ket{m} \bra{n} \rho_{(0,k)} \ket{n} \delta(E_k-(E_{m,k} - E_{n,k})),
\end{equation}
where $n$ and $m$ can be any of possible initial condition for a momentum $k$, the term $\bra{n} \rho_{0,k} \ket{n}$ gives the probability of finding the system in the state $(n,k)$ and the term $\bra{m} \rho_{k}(t) \ket{m}$ is the probability of finding the sytem in $(m,k)$ given the initial state $(n,k)$. 

Using Eq.\ref{probxchange}, the probability of A giving a particle for B and vice versa (for a momentum $k$) are
\begin{equation}\label{PAB}
P_{k,A \to B} (t) = n_{A,k}(1-n_{B,k})(1- e^{-\lambda t}cos(2g_k t))
\end{equation}
and
\begin{equation}\label{PBA}
P_{k,B \to A}(t) = n_{B,k}(1-n_{A,k})(1- e^{-\lambda t}cos(2g_k t))
\end{equation}
Thus, the fluctuation theorem for energy and matter of one mode will be
\begin{equation} \label{exfinal}
\frac{P_{A \to B}([\Delta E_A,\Delta N_A]_k)}{P_{B \to A}(-[\Delta E_A,\Delta N_A]_k)} =  e^{\Delta E_{A,k} F_H +\Delta N_{A,k} F_m},
\end{equation}
in which $F_H = \beta_B - \beta_A$ and $F_M =\beta_A \mu_A  -\beta_B \mu_B$ are the affinities related with energy and matter respectively, $\Delta N_{A,k} = -\Delta N_{B,k} = -1$ and $\Delta E_{A,k} = -\Delta E_{B,k} = -\epsilon_k$. 

As the modes are independents, the probabilities envolving each one is independent as well. So, the total probability of a total exchange $\Delta E_A$ and $\Delta N_A$ is basically the product of the probabilities envolving each individual mode. Therefore, the all-mode fluctuation theorem has the same shape of eq.(\ref{exfinal}), just taking away the $k$ indices.

Although the original formulations did not consider decoherence effects, we showed that their results may be obtained even if the system evolves under the influence of a dephasing.

\section{Conclusion} \label{conclusec}
The goal of this paper was to analyze  many possible aspects of the influence of a dephasing noise over a thermalization of a 1D bipartite fermionic lattice. 

We showed the role of this noise obtaining the equilibrium in some physical quantities. We gave the analytical solution for Boltzmann and Fermi-Dirac distribution in the fluxes analysis  Although quite different, Boltzmann may help us if we are dealing with semiconductors  and Fermi-Dirac for metals.

Moreover, we show that the main influence of dephasing is in  the arising of the second law of thermodynamics, an unfulfilled task from closed quantum systems approach. Finally, we showed that even in the presence of a dephasing noise,  the system obey a fluctuation theorem for energy and matter exchange.

As a final remark, it is important to comment that we used a phenomenological dephasing to do this work. A natural and possible sequence would be  a microscopic derivation of the evolution, given the interaction of the system with an environment. If a markovian evolution (with the weak coupling assumption) is considered, it  leads naturally to a equation of evolution which may be written in the Lindblad form. Therefore, even for  a microscopically obtained decoherence factor, the qualitative features of the evolution are the same presented in this work.
 
\section*{Acknowledgments}
I would like to thank CNPq, for the financial support. I also thank  Wallace  Teixeira, Ivan Medina and professor  Gabriel Landi, for the  fruitful discussions.      

\appendix

\section{The $\omega _{\nu}(x,y)$ function}  \label{omegafunction}
In contexts in which we use Boltzmann's distribution, integrals whose depend on powers of the mean occupation and/or powers of energy appear in many physical quantities. In this section we will compute a general case and define the $\omega _{\nu}(x,y)$ function, which will be very useful in the description of our quantities.

  Now, consider the following general integral:
\begin{equation}
\omega_{\nu}(x,y) = \frac{1}{\pi} \int_0^\pi \cos^\nu(z) e^{y \cos(z)}\cos(x \sin^2(z))\ud z. 
\end{equation} 
The first step is to expand the argument of the integral only in cosines dependences. So, using 
\begin{equation}
\cos(x) = \sum_{n=0}^\infty \frac{(-1)^n x^{2n}}{(2n)!} \quad \text{and} \quad e^x = \sum_{m=0}^\infty \frac{x^m}{m!}
\end{equation}
we obtain
\begin{eqnarray}
&\omega &_\nu(x,y)= \nonumber \\  & &\frac{1}{\pi}\sum_{n,m=0}^\infty \frac{(-1)^n x^{2n}}{(2n)!}\frac{y^m}{m!} \int_0^\pi \cos^{\nu + m}(z)(1-\cos^2(z))^{2n} \ud z. \nonumber \\
\end{eqnarray}
By a substitution of variables, 
\begin{equation}
\small
\omega _\nu(x,y)=\!\! \frac{1}{\pi}\!\sum_{n,m=0}^\infty \!\!\! \frac{(-1)^n x^{2n}}{(2n)!}\frac{y^m}{m!} \int_{-1}^1 u^{\nu + m}(1- u^2)^{2n -1/2} \ud u.
\normalsize
\end{equation}
When  $\nu + m$ is odd, the argument of the integral is odd as well and consequently, once the limits of integration are simmetric in relation to the origin, the integral is zero. On the other hand, if $\nu + m$ is even, this is a simple beta function. So, the general form of  $\omega _\nu(x,y)$ is 
\begin{equation}
\omega _\nu(x,y)=\!\! \frac{1}{\pi}\!\!\sum_{n,m=0}^\infty \!\!\! \frac{(-1)^n x^{2n}}{(2n)!}\frac{y^{2m+i}}{(2m+i)!} B \Big( 2n + \tfrac{1}{2}, \tfrac{\nu + 2m + 1 + i}{2}\Big),
\end{equation}  
in which $i=0$ when $\nu$ is even, $i=1$ when $\nu$ is odd and $B(n,m)$ are beta functions.


\section{Analytical solutions for particle and heat fluxes} \label{analytical}

As shown in section \ref{heatsec}, the quantities  $\mathcal{J_N}$ and $\mathcal{J_Q}$ may be straightfowardly derived by knowing  $\bar{N}$ and $\bar{E}$.   So, our task here is to derive these last functions.

 The analytical results for Boltzmann are labeled by $\bar{N}_B$ and $\bar{E}_B$. Analogously, for Fermi-Dirac, these quantities are labeled by $\bar{N}_{FD}$ and $\bar{E}_{FD}$.  For Boltzmann, using the $f_{abc}$, we obtain
\begin{eqnarray}
\bar{N}_B &=&   e^{\beta \mu} \Big(e^{-\lambda t} \omega_0(2 gt, 2 \beta) - I_0(2\beta)\Big)    \\
\bar{E}_B &=& -2 e^{\beta \mu}\Big(  e^{-\lambda t} \omega_1(2 gt, 2 \beta) - I_1(2\beta)\Big). 
\end{eqnarray} 

\begin{figure}[!h]	
\flushright
\includegraphics[width = 0.55 \textwidth ]{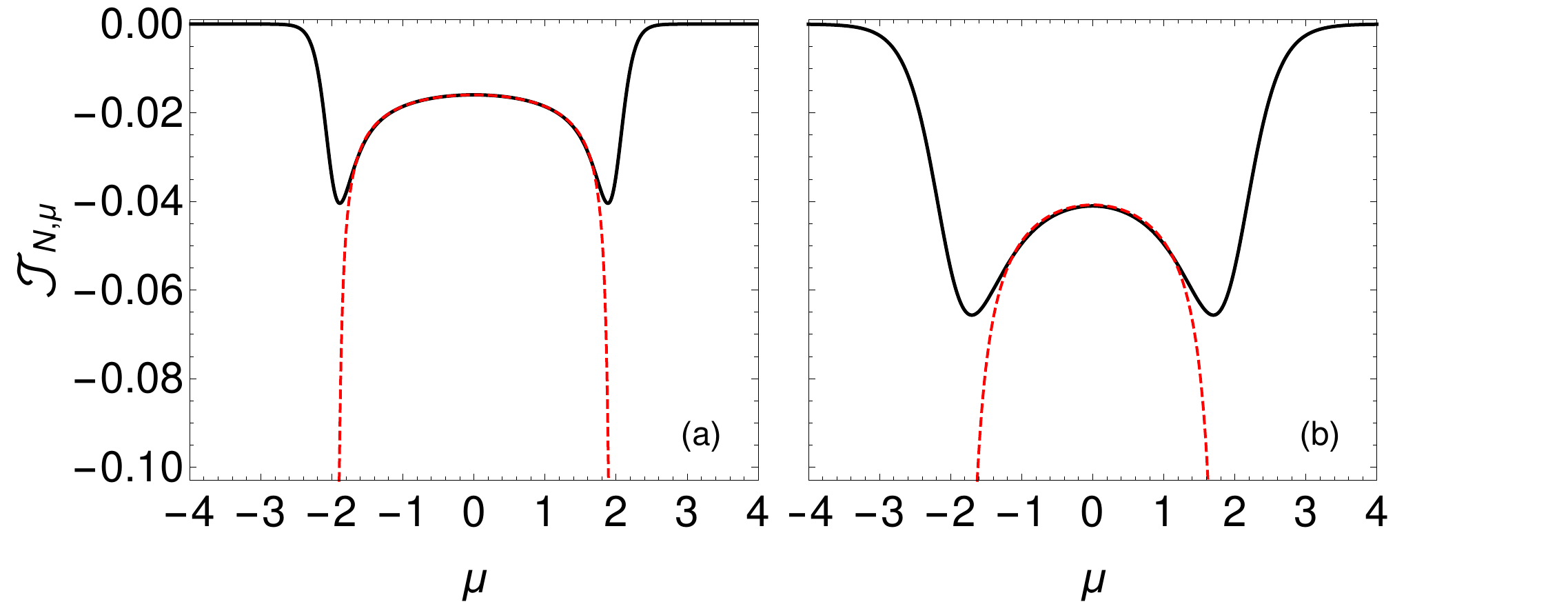}
\caption{\co : Onsager coefficients for $t \to \infty$ (equilibrium) as function of chemical potential in which  $\mu$ and $T$ are in units of $\alpha$. The solid line corresponds to the exact numerical result and the red dashed line, the analytical result. In figure (a) $T= 0.1$ and in figure (b) $T= 0.25$  }
\label{onsteste1}
\end{figure}

In  section \ref{heatsec} we discussed a simple way to analyze how good is to consider the Boltzmann distribution for those cases of interest.  On the other hand,   the low temperature approaximation for Fermi-Dirac statistics, through the Sommerfeld expansion, is not straightfowardly obtained. Even so, comparing the analytical and the numerical results, represented by Fig.\ref{onsteste1}, we obtain some restrictions. As soon as the temperature grows, the interval of chemical potential in which the approximation is valid becomes more reduced. In such way, the maximum temperature obtained was of order of $T \approx 0.8$, in which the approaximation is valid only for $\mu \approx 0$. A comparative of such behavior is shown in Fig\ref{onsteste1}. 
On the other hand, an analytical solution using Fermi-Dirac distribution can be obtained for small temperatures using Sommerfeld expansion. Using it, we obtain
\begin{widetext}
\begin{eqnarray}
\bar{N}_{FD} &=& 
\frac{1}{2\pi} \left\{ e^{-\lambda t} \left(\sum_0^\infty (-1)^n f_n(\mu,gt) \right) - \arccos\left( \frac{-\mu}{2} \right) + \frac{\pi^2 T_X^2}{6} \frac{\partial}{\partial \epsilon} \left[\frac{e^{-\lambda t}\cos \left( 2 gt \left(\frac{4-\epsilon^2}{4} \right) \right)-1}{\sqrt{4 - \epsilon^2}} \right]_{\epsilon = \mu_X} \right\}  
\end{eqnarray} 
\begin{eqnarray}
\bar{E}_{FD} &=&    \frac{e^{-\lambda t}}{\pi}\left(\sum_0^\infty (-1)^n h_n(\mu,gt) \right) +\frac{\sqrt{4-\mu^2}}{2 \pi}  +\!\! \frac{\pi T^2}{12}\frac{\partial}{\partial \epsilon}\left[ \frac{e^{-\lambda t}\epsilon \cos(g_\mu t)- 1}{\sqrt{4-\epsilon^2}} \right]_{\epsilon = \mu}  \nonumber \\
&&
\end{eqnarray}  
in which
	
\begin{equation}
f_{n}(\mu,gt) = \begin{cases} J_0(gt) \arccos\left(\frac{-\mu}{2} \right)  & \text{if} \quad n =0  \\
\frac{ \cos(gt) J_{2n}(gt) V_{4n} \left(\frac{-\mu}{2}\right)}{2n} - \frac{ \sin(gt) J_{2n-1}(gt)V_{4n -2} \left(\frac{-\mu}{2}\right)}{2n-1}  & \text{if} \quad n \neq 0 
\end{cases}
\end{equation}
and
\begin{equation}
h_{n}(\mu,gt) = \begin{cases} J_0(gt) \sqrt{1- \frac{\mu^2}{4}}   & \text{if} \quad n =0  \\
 \cos(gt) J_{2n}(gt) \left(\frac{V_{4n+1} \left(\frac{-\mu}{2}\right)}{4n+1} +\frac{V_{4n-1} \left(\frac{-\mu}{2}\right)}{4n-1} \right)+ \sin(gt) J_{2n-1}(gt) \left(\frac{V_{4n-1} \left(\frac{-\mu}{2}\right)}{4n-1} +\frac{V_{4n-3} \left(\frac{-\mu}{2}\right)}{4n-3} \right)  & \text{if} \quad n \neq 0 
\end{cases}
\end{equation}
Moreover, $g_\epsilon \equiv 2 g \big(1- \tfrac{\epsilon ^2}{4} \big)$ and $V_n(x)$,  are the third kind  Chebyshev functions.
\end{widetext}

\begin{figure}[!h]
\centering
\includegraphics[scale=0.65]{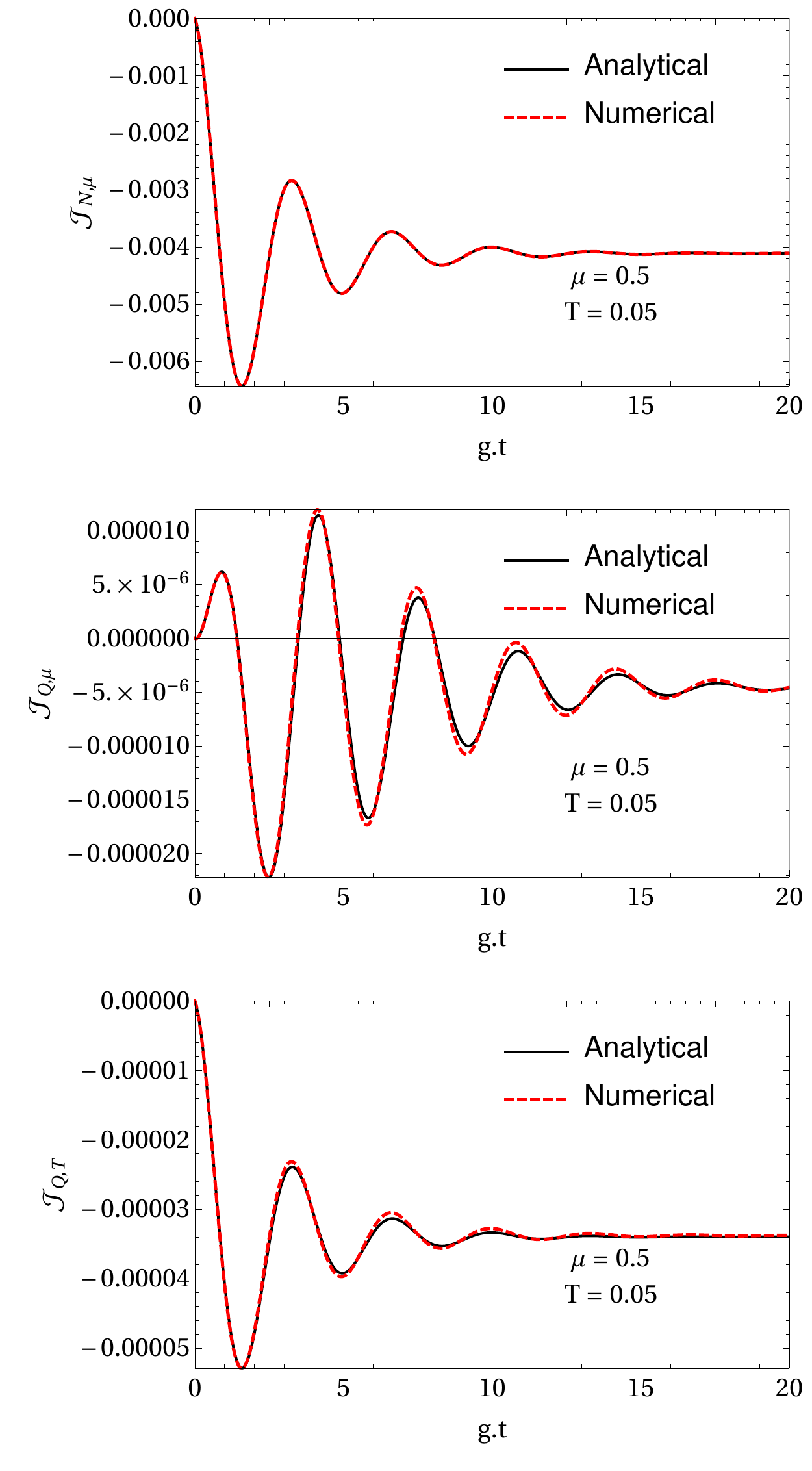}
\caption{\co: Test of validity for the analytical solutions. The solid black lines correspond to the analytical solutions and the red dashed lines correspond to the  exact results obtained numerically. The damping factor $\lambda = 0.35$ and $g =1$ were used here.}
\label{onsteste2}
\end{figure} 

In order to test the robustness of the solution, we plot the fluxes as shown  in Fig.(\ref{onsteste2}). Although it is possible to see slight deviations from the exact result, the analytical solution agree very well with the numerical one. Moreover, this difference tends to disappear as soon as $T \to 0$, completing the validity of the result.

\bibliographystyle{ieeetr}
\bibliography{references}

\begin{thebibliography}{10}

\bibitem{callen}
H.~B. Callen, {\em Thermodynamics}.
\newblock Wiley and Sons Inc., 1960.
\newblock chapter 16.

\bibitem{onsager1}
L.~Onsager, ``Reciprocal relations in irreversible processes. i.,'' {\em
  Physical review}, vol.~37, no.~4, p.~405, 1931.

\bibitem{onsager2}
L.~Onsager, ``Reciprocal relations in irreversible processes. ii.,'' {\em
  Physical review}, vol.~38, no.~12, p.~2265, 1931.

\bibitem{salinas1}
S.~Salinas, {\em Introdução à Física Estatística}.
\newblock EdUsp, 1999.
\newblock Capítulo 1.

\bibitem{crooks}
G.~E. Crooks, ``Nonequilibrium measurements of free energy differences for
  microscopically reversible markovian systems,'' {\em Journal of Statistical
  Physics}, vol.~90, no.~5-6, pp.~1481--1487, 1998.

\bibitem{jarzynski}
C.~Jarzynski and D.~K. W{\'o}jcik, ``Classical and quantum fluctuation theorems
  for heat exchange,'' {\em Physical review letters}, vol.~92, no.~23,
  p.~230602, 2004.

\bibitem{landi2016}
G.~T. Landi and D.~Karevski, ``Fluctuations of the heat exchanged between two
  quantum spin chains,'' {\em Physical Review E}, vol.~93, no.~3, p.~032122,
  2016.

\bibitem{saito}
K.~Saito and A.~Dhar, ``Fluctuation theorem in quantum heat conduction,'' {\em
  Phys. Rev. Lett.}, vol.~99, p.~180601, Oct 2007.

\bibitem{zon}
R.~van Zon, S.~Ciliberto, and E.~G.~D. Cohen, ``Power and heat fluctuation
  theorems for electric circuits,'' {\em Phys. Rev. Lett.}, vol.~92, p.~130601,
  Mar 2004.

\bibitem{ciliberto}
J.~Gomez-Solano, A.~Petrosyan, and S.~Ciliberto, ``Heat fluctuations in a
  nonequilibrium bath,'' {\em Physical review letters}, vol.~106, no.~20,
  p.~200602, 2011.

\bibitem{ciliberto2}
N.~Garnier and S.~Ciliberto, ``Nonequilibrium fluctuations in a resistor,''
  {\em Phys. Rev. E}, vol.~71, p.~060101, Jun 2005.

\bibitem{collin}
D.~Collin, F.~Ritort, C.~Jarzynski, S.~B. Smith, I.~Tinoco, and C.~Bustamante,
  ``Verification of the crooks fluctuation theorem and recovery of rna folding
  free energies,'' {\em Nature}, vol.~437, no.~7056, pp.~231--234, 2005.

\bibitem{serra}
T.~B. Batalh\~ao, A.~M. Souza, L.~Mazzola, R.~Auccaise, R.~S. Sarthour, I.~S.
  Oliveira, J.~Goold, G.~De~Chiara, M.~Paternostro, and R.~M. Serra,
  ``Experimental reconstruction of work distribution and study of fluctuation
  relations in a closed quantum system,'' {\em Phys. Rev. Lett.}, vol.~113,
  p.~140601, Oct 2014.

\bibitem{horo}
M.~Horodecki and J.~Oppenheim, ``Fundamental limitations for quantum and
  nanoscale thermodynamics,'' {\em Nature communications}, vol.~4, 2013.

\bibitem{kosloff}
R.~Kosloff, ``Quantum thermodynamics: A dynamical viewpoint,'' {\em Entropy},
  vol.~15, no.~6, pp.~2100--2128, 2013.

\bibitem{vinjanampathy}
S.~Vinjanampathy and J.~Anders, ``Quantum thermodynamics,'' {\em Contemporary
  Physics}, vol.~57, no.~4, pp.~545--579, 2016.

\bibitem{pekola2015}
J.~P. Pekola, ``Towards quantum thermodynamics in electronic circuits,'' {\em
  Nature Physics}, vol.~11, no.~2, pp.~118--123, 2015.

\bibitem{prance}
J.~Prance, C.~Smith, J.~Griffiths, S.~Chorley, D.~Anderson, G.~Jones,
  I.~Farrer, and D.~Ritchie, ``Electronic refrigeration of a two-dimensional
  electron gas,'' {\em Physical review letters}, vol.~102, no.~14, p.~146602,
  2009.

\bibitem{jezouin}
S.~Jezouin, F.~Parmentier, A.~Anthore, U.~Gennser, A.~Cavanna, Y.~Jin, and
  F.~Pierre, ``Quantum limit of heat flow across a single electronic channel,''
  {\em Science}, vol.~342, no.~6158, pp.~601--604, 2013.

\bibitem{ohdissi}
I.~Senitzky, ``Dissipation in quantum mechanics. the harmonic oscillator,''
  {\em Physical Review}, vol.~119, no.~2, p.~670, 1960.

\bibitem{ohlutz}
S.~Deffner and E.~Lutz, ``Nonequilibrium work distribution of a quantum
  harmonic oscillator,'' {\em Phys. Rev. E}, vol.~77, p.~021128, Feb 2008.

\bibitem{anichain}
A.~Kl{\"u}mper, ``Thermodynamics of the anisotropic spin-1/2 heisenberg chain
  and related quantum chains,'' {\em Zeitschrift f{\"u}r Physik B Condensed
  Matter}, vol.~91, no.~4, pp.~507--519, 1993.

\bibitem{vznidarivc}
M.~{\v{Z}}nidari{\v{c}}, ``Exact solution for a diffusive nonequilibrium steady
  state of an open quantum chain,'' {\em Journal of Statistical Mechanics:
  Theory and Experiment}, vol.~2010, no.~05, p.~L05002, 2010.

\bibitem{prosen2010}
T.~Prosen and M.~{\v{Z}}nidari{\v{c}}, ``Long-range order in nonequilibrium
  interacting quantum spin chains,'' {\em Physical review letters}, vol.~105,
  no.~6, p.~060603, 2010.

\bibitem{kaufman2016}
A.~M. Kaufman, M.~E. Tai, A.~Lukin, M.~Rispoli, R.~Schittko, P.~M. Preiss, and
  M.~Greiner, ``Quantum thermalization through entanglement in an isolated
  many-body system,'' {\em Science}, vol.~353, no.~6301, pp.~794--800, 2016.

\bibitem{medina}
I.~Medina and F.~Semiao, ``Pulse engineering for population control under
  dephasing and dissipation,'' {\em Physical Review A}, vol.~100, no.~1,
  p.~012103, 2019.

\bibitem{rieder}
Z.~Rieder, J.~Lebowitz, and E.~Lieb, ``Properties of a harmonic crystal in a
  stationary nonequilibrium state,'' {\em Journal of Mathematical Physics},
  vol.~8, no.~5, pp.~1073--1078, 1967.

\bibitem{stefano}
S.~Lepri, R.~Livi, and A.~Politi, ``Heat conduction in chains of nonlinear
  oscillators,'' {\em Phys. Rev. Lett.}, vol.~78, pp.~1896--1899, Mar 1997.

\bibitem{narayan}
O.~Narayan and S.~Ramaswamy, ``Anomalous heat conduction in one-dimensional
  momentum-conserving systems,'' {\em Phys. Rev. Lett.}, vol.~89, p.~200601,
  Oct 2002.

\bibitem{landi2013}
G.~T. Landi and M.~J. de~Oliveira, ``Fourier's law from a chain of coupled
  anharmonic oscillators under energy-conserving noise,'' {\em Physical Review
  E}, vol.~87, no.~5, p.~052126, 2013.

\bibitem{chuang}
M.~A. Nielsen and I.~L. Chuang, {\em Quantum computation and Quantum
  Information}.
\newblock Cambridge University Press, 2000.

\bibitem{zeng}
X.~C. Bei~Zeng, {\em Quantum Information Meets Quantum Matter}.
\newblock Springer, 2016.
\newblock Section 1.3.

\end{thebibliography}

\end{document}